\documentclass[twocolumn,showpacs,prb]{revtex4}

\usepackage{amsmath}
\usepackage{amsmath}
\usepackage{amssymb}
\usepackage{graphicx}

\newcommand{\rvec}{\mathbf{r}}
\newcommand{\Rvec}{\mathbf{R}}

\begin{document}

\title{All-electron density functional theory and time-dependent
  density functional theory with high-order finite elements}
\author{Lauri Lehtovaara$^1$}
\author{Ville Havu$^{1,2}$}
\author{Martti Puska$^1$}
\affiliation{
$^1$Department of Applied Physics / COMP
and
$^2$Institute of Mathematics,
Helsinki University of Technology,
P.O.Box 110, FIN-02015 TKK, Finland
}

\begin{abstract}

We present for static density functional theory and time-dependent
density functional theory calculations an all-electron method which
employs high-order hierarchical finite-element bases. Our mesh
generation scheme, in which structured atomic meshes are merged to an
unstructured molecular mesh, allows a highly nonuniform discretization
of the space. Thus it is possible to represent the core and valence
states using the same discretization scheme, {\em i.e.}, no
pseudopotentials or similar treatments are required. The nonuniform
discretization also allows the use of large simulation cells, and
therefore avoids any boundary effects.

\end{abstract}

\maketitle

\section{Introduction}

The density functional theory (DFT) has become the workhorse in
electronic structure theory \cite{kohn_nobel_99}. Its success derives
from the ability to
produce accurate results with reasonable computational effort. Instead
of solving for the many-body wavefunction it relies on Hohenberg-Kohn
theorem \cite{hohenberg_kohn_64}
which states that all ground-state properties -- most importantly the
total energy -- are functionals of the electron density. Actually,
the total energy functional is not exactly know, but there exists several
approximations, the accuracy of which can be systematically improved
according to the demands of the applications in question 
\cite{parr_yang_dft_book, dft_primer}. 
The most important issue is that the number of dimensions in a
problem does not depend on the number of electrons, and thus DFT
scales much better than many-body wavefunction methods, up to hundreds or
thousands of atoms on the present supercomputers. 

The DFT is bound to the ground-state properties
and cannot be used to explore excited states. This drawback
can be overcome by using the time-dependent DFT
(TDDFT) \cite{elliott_tddft_review_2008}. It is based on the
Runge-Gross theorem \cite{runge_gross_84}, which states that
(physically) different external potentials ({\em e.g.}, those due 
to laser fields) lead to different time-evolutions of the density. The present
functionals for TDDFT are known to be unable to describe certain
phenomena, such as charge transfer excitations. However, in recent
years it has been successfully applied to description several other 
problems, for example, the optical absorption spectra of a broad variety of
systems, the nonlinear optical response ({\em e.g.}, harmonic generation)
of atoms and molecules, and coherent control of molecules by laser
fields \cite{elliott_tddft_review_2008}.

For numerical solution, the partial differential equations arising
from DFT and TDDFT must be discretized in space. In the present-day codes,
the most popular choices are atomic orbital bases \cite{Gaussian03,
  Siesta, FHIaims}, planewaves \cite{VASP, AbInit}, and uniform
real-space grids \cite{GPAW, Octopus}. In the atomic orbital
bases the solution is represented as a linear
combination of atomic solutions, which can be
accurate ({\em e.g.}, numerical atomic orbitals \cite{FHIaims}) or approximate
({\em e.g.}, Gaussians \cite{Gaussian03}). These bases are widely used
and can be very fast and efficient. However, the atomic orbital bases
are sensitive to the type of the problem
in the sense that an efficient discretization for the ground state
properties is not well suited for the calculation of optical absorption
spectra. In particular, when the solution is not representable as
slightly perturbed atomic solutions the atomic orbital bases become
unfavourable. For example, this can happen in the case of nonlinear
time-dependent phenomena.

The planewave bases and uniform real-space grids ({\em i.e.}, the
finite-difference method) are both uniform
discretizations of the space and closely related to each other through
the Fourier transform. These discretizations are not dependent on the type
of the problem, but they require a large number of degrees of
freedom. Especially, the core regions around nuclei, where solutions
have very sharp features, cannot be represented well by uniform
discretization, but pseudopotentials \cite{fermi_pp_34,
  phillips_pp_59, troullier_martins_pp_91} or similar treatments
({\em e.g.}, projector-augmented wave method \cite{bloch_paw_94}) must
be employed. The pseudopotentials lead to
additional parameters and may be hard to construct accurately for
certain types of atoms, {\em e.g.}, transition metals. Another drawback
in uniform discretizations is their inability to adapt to the underlying
geometry of the atoms. For example, in sparse matter interstitial
regions should require much less degrees of freedom than regions near
atoms. This is also the case in simulations of nonlinear
time-dependent phenomena, where the distant regions in space should
still be accounted for but the solution is smooth in this region so
that the discretization can be coarse.

The finite element basis \cite{turner_fem_56, szabo_babuska_fea_book} is
a linear combination of continuous, piece-wise
polynomials and provides a nonuniform real-space discretization
of the space. It inherits the good properties of the real-space
methods, such as, flexible boundary conditions and efficient
parallelization via domain decomposition, while still allowing
nonuniform discretization of the space. In this paper,
we use high-order hierarchical
finite elements, which i) provide a better rate of convergence than
low-order elements, and ii) result in better conditioned systems of
linear equations than the nodal-based elements of the same order. As
finite elements can adapt to the local feature size, they can be used
to describe solutions of core and valence electrons equally well. And
naturally, they are adaptable to
the geometry of the system to avoid overdiscretisation. The finite
element basis is also variational like planewaves and atomic bases
which is not the case for finite-difference discretizations. The
finite element basis is extensively used in civil and mechanical
engineering, and in many fields it has surpassed finite-difference
methods \cite{bathe_fem_book}. There exists several earlier
implementations of the finite-element methodology to electronic
structure calculations \cite{pask_sterne_review_05, zhang_2008,
  fattebert_2007, tsuchida_95, white_89, yu_h2_fem_94, batcho_98,
  batcho_00, levin_85}.
However, none of these uses high-order hierarchical elements on
unstructured meshes or apply the method to TDDFT. The closest work to
our approach is the spectral element method implementation of Batcho
\cite{batcho_00}. The spectral element method uses high-order tensor
product bases, which enable fast evaluation of matrix elements and
provide good convergence rates. However, the element types are
restricted to brick ({\em i.e.}, parallelepiped) elements and mapped
brick elements ({\em i.e.}, coordinate transformations of brick
elements). 

The rest of the paper is structured as follows. In the next section,
we briefly review the DFT, linear response TDDFT, and finite-element
method. We also describe our mesh generation algorithm. In
the section \ref{sec:results}, we show several examples of applying
our DFT and
linear-response TDDFT method to small molecules (CO, Na$_2$,
C$_6$H$_6$) and discuss the convergence of the method. In the final
section, we draw the conclusions and set directions for future
research.

\section{Theory}\label{sec:theory}

\subsection{Density functional theory}

In the density functional theory, the total energy $E[n(\rvec)]$ is a
functional of the electron density $n(\rvec)$, and the ground state of
the system is found by minimizing
it. However, the functional is not known in general and must be
approximated. This is usually done by employing the Kohn-Sham
\cite{kohn_sham_65} scheme where the functional is divided into four
parts:
\begin{equation}
  E[n] = T_s[n] + \int d^3 r n(\rvec) v_{ext}(\rvec) + U[n] + E_{xc}[n],
\end{equation}
where $T_s[n]$ is kinetic energy of the non-interacting electron
system with density $n(\rvec)$,
$\int d^3 r n(\rvec) v_{ext}(\rvec)$ is the interaction energy with an
external field (usually, that due to the ions), $U[n]$ is the mean
electron-electron repulsion energy (Hartree energy), and $E_{xc}[n]$
is the electron exchange-correlation energy functional. The three
first parts are known but the last one, the exchange-correlation
functional, is not, and the quality of its approximation is the key to
accurate results. The Kohn-Sham scheme uses a set of orthonormal
auxiliary functions $\psi_k(\rvec)$, {\em i.e.}, the Kohn-Sham
orbitals, which satisfy 
\begin{equation}
n(\rvec) = \sum_{k=1}^{N_{states}} f_k |\psi_k(\rvec)|^2,
\end{equation}
where $f_k$ are the occupation numbers, and $N_{states}$ is the number
of occupied Kohn-Sham orbitals. By taking the functional
derivative of the energy functional with respect to these functions,
we obtain the Kohn-Sham equations:
\begin{equation}\label{eq:Kohn_Sham}
  \hat{H}_{KS} \psi_k(\rvec) =
  \left(- \frac{\hbar^2}{2m_e} \nabla^2
    + v_{eff}(\rvec) \right) \psi_k(\rvec) = \epsilon_k \psi_k(\rvec),
\end{equation}
where
\begin{equation}
v_{eff}(\rvec) = v_H[n](\rvec) + v_{xc}[n](\rvec) + v_{ext}(\rvec)
\end{equation}
is the effective potential, and
\begin{equation}
v_H[n](\rvec) = \frac{e^2}{4\pi\varepsilon_0}
\int d^3 r' \frac{n(\rvec')}{|\rvec-\rvec'|}
\end{equation}
is the Hartree potential. Furthermore, $v_{xc}[n](\rvec)$ is the
exchange-correlation potential, and $v_{ext}(\rvec)$ is the external
potential, which is usually a sum of electron-nucleus interactions,
{\em i.e.},
\begin{equation}
v_{ext}(\rvec) = \frac{-e^2}{4\pi\varepsilon_0}
\sum_{a=1}^{N_{nuclei}} \frac{Z_a}{|\rvec - \rvec_a|},
\end{equation}
where $Z_a$ is the atomic number and $\rvec_a$ is the position of the
nucleus $a$. $N_{nuclei}$ is the number of nuclei in the system.
In the three dimensional space $\mathbb{R}^3$, the Hartree potential can
be rewritten as the solution of the Poisson equation
\begin{equation}\label{eq:Poisson}
  \nabla^2 v_H(\rvec) =  -4\pi \frac{e^2}{4\pi\varepsilon_0} n(\rvec),
\end{equation}
where the boundary condition for isolated systems is $v_H \rightarrow
0$ when $|\rvec| \rightarrow \infty$. (Also periodic and other
boundary conditions are possible but are not discussed in this paper.)

As the Hartree potential, the density and thus the Kohn-Sham
wavefunctions vanish at the infinity (or in practice at the boundary
$\partial \Omega$ of the computational domain $\Omega$), the above
Eqs.~\eqref{eq:Kohn_Sham} and \eqref{eq:Poisson} can be cast into the
weak variational formulation using integration by parts, {\em i.e.},
\begin{multline}\label{eq:Kohn_Sham_weak}
\langle \Phi | \hat{H}_{KS} | \psi_k \rangle = \\
\int_{\mathbb{R}^3} \Phi(\rvec) \left( \frac{-\hbar^2}{2m_e} \nabla^2 +
v_{eff}(\rvec) \right) \psi_k(\rvec) d^3 r
\\
=
\int_{\mathbb{R}^3} \left(
\frac{\hbar^2}{2m_e} \nabla \Phi(\rvec) \cdot \nabla \psi_k(\rvec)
\right.
\\
\left. \frac{}{} + \Phi(\rvec) v_{eff}(\rvec) \psi_k(\rvec) \right) d^3 r,
\end{multline}
and
\begin{multline}\label{eq:Poisson_weak}
\langle \Phi | \nabla^2 | v_H \rangle =
\int_{\mathbb{R}^3} \Phi(\rvec) \nabla^2 v_H(\rvec) d^3 r =
\\
- \int_{\mathbb{R}^3} \nabla \Phi(\rvec) \cdot \nabla v_H(\rvec) d^3 r,
\end{multline}
where $\Phi(\rvec)$ is a test function which has a square
integrable gradient $\nabla \Phi(\rvec)$. The weak formulation is
required by the finite element method, and in practice, $\Phi(\rvec)$
will be a finite element basis function (in the so-called Ritz-Galerkin
method \cite{brenner_scott_book}, see Eq.~\eqref{eq:Kohn_Sham_fem}).

As the Hartree potential for charged systems decays slowly as  $r^{-1}$,
we have applied counter charges to neutralize the density. The
counter charges are added to the electronic density $n(r)$ in
Eq.~\eqref{eq:Poisson_weak} and are then cancelled in
Eq.~\eqref{eq:Kohn_Sham_weak} by the corresponding analytically
calculated potential. This provides the $r^{-2}$ decay of the Hartree
potential, which is sufficient for our purposes. However, if required,
higher order ({\em e.g.}, dipole and quadrupole) corrections can be
applied as well \cite{batcho_00}.

\subsection{Linear response time-dependent DFT}

In the time-dependent DFT, there exist no variational principle, but
the quantum mechanical action
\begin{equation}
  A[\psi] = \int_{t_0}^{t_1} dt \langle \psi(t) | i \hbar
  \frac{\partial}{\partial t} - \hat{H}(t) | \psi(t) \rangle
\end{equation}
provides an analogous quantity to the total energy of the ground-state
DFT. The time-dependent Kohn-Sham Schr\"odinger equation reads as
\begin{equation}
  i \hbar \frac{\partial}{\partial t} \psi_k(\rvec,t) =
  \left( - \frac{\hbar^2}{2m_e} \nabla^2 + v_{eff}[n](\rvec,t) \right)
  \psi_k(\rvec,t).
\end{equation}
This equation is an initial value problem and can be solved using
a time-propagation scheme \cite{yabana_tdlda_96}. However, if the
external perturbation is small, the density response of the system can
be written as a series
\begin{equation}
  n(\rvec,\omega) = n^{(0)}(\rvec) + n^{(1)}(\rvec,\omega)
  + n^{(2)}(\rvec,\omega) + \dots,
\end{equation}
with the linear response term
\begin{equation} \label{eq:density_response}
  n^{(1)}(\rvec,\omega) = \int d^3 r' \chi(\rvec, \rvec',\omega)
  v^{(1)}(\rvec',\omega).
\end{equation}
Above, $\chi$ is the linear response function and $v^{(1)}$ is the external
perturbation ({\em e.g.} a laser field). The transitions can be
found by finding the poles of the response function $\chi(\rvec,
\rvec',\omega)$. However, if we are interested only in the excitation
energies and corresponding oscillator strengths, we can use the
so-called Casida method. He showed that the problem can be solved as
an eigenvalue equation \cite{casida_lrtddft_95, casida_lrtddft_96}
\begin{multline}
  \sum_{j'k'}
  \left[ \delta_{jk} \delta_{j'k'} \epsilon_{jk}^2 +
    2 \sqrt{f_{kj} \epsilon_{jk}
      f_{k'j'} \epsilon_{j'k'} }
    K_{jk,j'k'} \right] \\
  \times
  \gamma_{j'k'}
  = \Omega^2 \gamma_{jk},
\end{multline}
where $f_{kj} = f_{k} - f_{j}$, $\epsilon_{jk} = \epsilon_{j} -
\epsilon_{k}$, and
the coupling matrix
\begin{multline}\label{eq:LR_integral}
  K_{jk,j'k'}(\omega) = \\
  \int d^3 \rvec \int d^3 \rvec'
  \psi_j^*(\rvec) \psi_k(\rvec) \psi_{j'}(\rvec') \psi_{k'}^*(\rvec')
  \\
  \times
  \left[
    \frac{e^2}{4\pi\varepsilon_0} \frac{1}{|\rvec-\rvec'|}
      \frac{}{} + f_{xc}(\rvec,\rvec',\omega)
  \right].
\end{multline}
Moreover,
\begin{equation}
  f_{xc}(\rvec \omega,\rvec' \omega') =
  \frac{\delta v_{xc}(\rvec,\omega)}{\delta n(\rvec',\omega')}
\end{equation}
is the exchange-correlation kernel. The oscillator strengths are then
\begin{equation}
  \tilde{f}^{(m)}_{x/y/z} = \frac{2m}{\hbar^2 e^2}
  \left| \sum_{jk}^{f_k>f_j} (\mu_{jk})_{x/y/z}
    \sqrt{(f_k - f_j) (\epsilon_j - \epsilon_k)} \gamma^{(m)}_{jk} \right|^2,
\end{equation}
where $(\mu_{jk})_{x/y/z}$ is the $x/y/z$ component of the dipole moment
vector between the Kohn-Sham states $k$ and $j$, and the index $(m)$
refers to the $m^{th}$ transition.

\subsubsection{Confinement potential}

The linear response Kohn-Sham equations use the Kohn-Sham states as a
basis. Above the ionization limit of the system, the spectrum becomes
continuous causing numerical problems. The eigenvalues of the
discretized problem bunch together just above the ionization limit.
For a practical calculation this is not desirable because
certain transitions have very many different contributions due to the  
eigenstates in the Kohn-Sham continuum and the importance of most 
of them is minor because the states have a relatively small amplitude near 
the molecule.

To spread the eigenvalue spectrum above the ionization limit, and to 
increase the relative importance of the relevant unoccupied
states, we use a modified Kohn-Sham basis $\{ \tilde{\psi}_k(\rvec) \}$.
The basis is constructed by applying an
auxiliary confinement potential in the ground-state calculation. The
choice of the potential is in principle arbitrary, but in order to
fill the above requirements, we have chosen the form
\begin{equation}
    v_{conf}(\rvec) =
    \left\{
      \begin{aligned}
        \frac{1}{2} k_c |r_{\min}(\rvec) - R_c|^2,
        &\quad \text{if } r_{min}(\rvec) > R_c \\
        0, & \quad \text{otherwise},
      \end{aligned}
    \right.
\end{equation}
where $r_{\min}(\rvec) = \min_{R_a} |r - R_a|$ is the distance to the
closest atom, and $k_c$ and $R_c$ are parameters to be chosen. Thus,
the auxiliary potential is zero close to the atoms but becomes
gradually more repulsive further away. Far away from the system, the
auxiliary potential is a spherically symmetric harmonic
potential. Now, all states are bound.

After the ground-state calculation with the auxiliary confinement
potential the resulting Kohn-Sham states $\{ \tilde{\psi}_k(\rvec) \}$
are taken as the new basis, the auxiliary confinement potential is
removed, and the ground-state calculation is repeated in the new
basis. Finally, the linear response calculation is carried out in the
new basis.

Introducing the auxiliary confinement potential allows us to balance
between the number of unoccupied states and the quality of the low
energy part of the spectrum. We want to stress out that this is purely
a mathematical trick in order to alter the basis of the linear
response calculation in such a way that the low energy transitions
converge more quickly. The physics is not altered. The calculated
linear response spectrum with and without an auxiliary confinement
potential should give the same result when all the Kohn-Sham states
(occupied and unoccupied) are used as they span the same original
finite element space $V_h$. Also, as the confinement potential
determines the linear response basis, the final result of a converged
calculation is independent of the original basis where the Kohn-Sham
states were solved, {\em e.g.}, converged atomic orbital and
real-space calculations should give the same result.

The choice of the parameters $R_c$ and $k_c$ is not an obvious task
and some testing is required to find appropriate values. However, the
testing can be done as a linear problem by fixing the density, because
the confinement should not change the ground-state.

\subsection{Finite-Element Discretization}

In the finite-element method the computational domain $\Omega$ is
divided into small, polyhedral regions called elements. This division
is denoted by $\mathcal{T}_h$. For our purposes it is sufficient to
use tetrahedra. Other popular choices are hexahedra, pyramids and
prisms. The division of $\Omega$ is handled by an external
mesh-generator that can either i) generate the mesh for a given
geometry or ii) calculate the 
Delaunay tetrahedralization of a given set of points. We have chosen
the latter option and the points for the mesh are generated as
specified in Section~\ref{mesh_generation}.

Once the division of the domain $\Omega$ is complete the space of
approximation, $V_h$, can be defined. For the finite-element method
this is taken to be continuous, piecewise polynomial functions, i.e.
\begin{equation}\label{fe_space}
 V_h = \{ v_h \in C(\Omega) \, | \, (v_h)_{|K} \in \Pi_p \} \quad
 \forall K \in \mathcal{T}_h
\end{equation}
where $K$ is an element, $\Pi_p$ denotes polynomials of order $p$,
$h$ refers to the size of the elements in the mesh, and $C(\Omega)$
refers to continuous functions in the domain. In
general, the order $p$ can vary from one element to another as long as
the continuity condition $v_h \in C(\Omega)$ is respected but in our
calculations we choose to keep $p$ fixed throughout the mesh. The
value of $p$ decides if the method is considered to be of high-order
and the usual requirement is $p > 3$ for a high-order method. Also, if
the convergence is obtained via increasing the order of polynomials
rather than refining the mesh the method is called the $p$-method.
The mesh refinement approach gives an $h$-method and combining these
approaches leads to an $hp$-method \cite{schwab_hp_book_98}.

\begin{figure}[ht]
  \includegraphics[width=3.in]{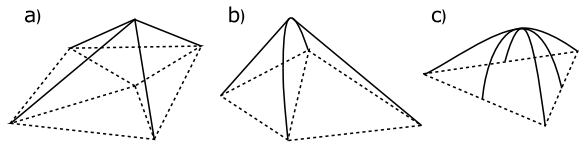}
  \caption{Schematic view of finite element basis functions in 2D: a) vertex,
    b) edge, and c) bubble basis functions}
  \label{fig:fem_basis_funcs}
\end{figure}

Next, a basis for the space $V_h$ must be chosen. The canonical way
for the high-order method is to divide the local basis functions of a
single element into four disjoint sets: nodal functions, edge
functions, face functions, and bubble functions. The nodal functions
are first order polynomials that have a value one at one of the
vertices and zero at others. The edge functions are polynomials up-to
an order $p$ and they are non-zero only on one of the edges of the
element. The face functions are similar to the edge functions but they
are in correspondence with the faces of the element. Finally, the
bubble functions are zero on all the vertices, edges and faces of the
element but non-zero inside the element. The actual basis functions
are generated using products of one-dimensional integrated Legendre
polynomials over the interval $[-1,1]$. Note that due to the
continuity requirements the basis functions actually extend over
several elements that share the same geometrical feature (see
Fig. \ref{fig:fem_basis_funcs}).

In practise, the basis functions for an element $K$ in the mesh are
generated using a reference element, $\hat{K}$, and (affine) mappings
$F: \hat{K} \to K$. Then the basis functions on an element $K$ can be
written as images of the basis functions on the reference element,
{\em i.e.},
\begin{equation}
 \varphi (\mathbf{r}) = \hat{\varphi} (F^{-1}(\mathbf{r})),
\end{equation}
reducing the programming effort to $\hat{K}$.

Once the basis $\{ \phi_j \}_{j=1}^{N_b}$ for the space $V_h$ is ready
for use an approximation to the Kohn-Sham orbitals can be looked for
in the form $\psi_k (\mathbf{r}) = \sum_{j=1}^{N_b} c_j^k \phi_j
(\mathbf{r})$. There are many ways to find the coefficients $c_i$ but
in the finite-element method the variational approach is used. This
leads to an equation for the state $k$
\begin{equation}\label{eq:Kohn_Sham_fem}
 \sum_{j=1}^{N_b} \langle \phi_i | \hat{H}_{KS} | \phi_j \rangle c_j^k
 = \epsilon_k \sum_{j=1}^{N_b} \langle \phi_i | \phi_j \rangle c_j^k,
 \quad i=1, \ldots, N_b,
\end{equation}
that reads in matrix form as
\begin{equation}
 Hc^k = \epsilon_k S c^k,
\end{equation}
where
\begin{equation}
 H_{ij} = \langle \phi_i | \hat{H}_{KS} | \phi_j \rangle,
 \quad S_{ij} = \langle \phi_i | \phi_j \rangle 
 = \int_{\mathbb{R}^3} \phi_i(\mathbf{r})\phi_j(\mathbf{r}) \,
 d\mathbf{r}.
\end{equation}
A few observations are in order. First, since the finite-element basis
functions are strictly localized in space the matrices $H$ and $S$ are
sparse. This not only allows for but actually dictates the use of
sparse matrix technologies. Second, if the domain $\Omega$ is large
enough so that selecting the zero boundary conditions on $\partial
\Omega$ is justified the variational formulation \eqref{eq:Kohn_Sham_fem}
holds and consequently the matrix $H$ is also symmetric. In this case
the fact that the basis functions $\phi_i$ don't have continuous
derivatives across the element borders is not an obstacle since in
\eqref{eq:Kohn_Sham_fem} only a square integrable gradient is required
for the basis functions (see Eq. \eqref{eq:Kohn_Sham_weak}).

\subsubsection{Mesh generation}\label{mesh_generation}

\begin{figure}[ht]
  \includegraphics[width=2.4in]{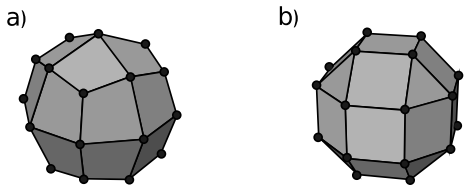}
  \caption{Polyhedra used in atomic meshes of a) deltoidal
    icositetrahedron and b) rhombicuboctahedron}
  \label{fig:polyhedra}
\end{figure}

The mesh is generated by merging structured atomic meshes to a
molecular mesh. The nodes of atomic meshes consist of layers of
vertices of polyhedra. The radius of the layer $r_k$ is changed
as $r_k = q^k r_0$ with $r_0$ and $q$ as parameters, and
$k \subset \mathbb{Z}$ ($-n \leq k \leq m; \, n,m \subset \mathbb{N}$).
The choice of polyhedra is arbitrary, but they should provide
tetrahedra of good quality (our quality requirements are explained
below in this section). We have chosen to use
deltoidal icositetrahedron and its dual, rhombicuboctahedron, both
shown in Fig. \ref{fig:polyhedra}.

The zeroth layer is chosen relative to the size of the highest occupied
atomic orbital $r_0 = (2I)^{-1/2}/4$, where $I$ is the first ionization
energy. The layers with negative indices are created until the radius
of the layer is of the order of the lowest state $r_{k_{\min}} <
Z_a^{-1}/128$. The
factors $\frac{1}{4}$ and $\frac{1}{128}$ are somewhat arbitrary at the
moment, but are sufficient for systems under study. If necessary one
extra layer is added, as the last layer should be deltoidal
icositetrahedron to ensure good quality of the elements around the
nuclei. The inner part of the mesh is finalized by adding one node to
the nucleus $\Rvec_a$.

The nodes of the layers with positive indices are added only if the
node is inside the atomic mesh region, {\em i.e.}, not in the
molecular mesh region. The node of atom $a$ is in the molecular region
if
\begin{equation}
g_{ab} |\Rvec_b - \Rvec_a| / |\rvec - \Rvec_a|
- \frac{\rvec - \Rvec_a}{|\rvec - \Rvec_a|}
\cdot
\frac{\Rvec_b - \Rvec_a}{|\Rvec_b - \Rvec_a|}
< \beta (q-1)
\end{equation}
for all other nuclei $b$, where $g_{ab} = r^a_0 / ( r^a_0 + r^b_0 )$
are the relative sizes with respect to the other nuclei, and $\beta$
is chosen to be $\frac{1}{3}$. In practice, this procedure
creates an empty space between atoms, which reaches closer to smaller
atoms than larger ones, and its thickness is proportional to the distance
between the closest pair of atoms. For each pair of atoms the atomic 
regions are inside two halves of an elliptical hyperboloid.

\begin{figure}[ht]
  \includegraphics[width=2.5in]{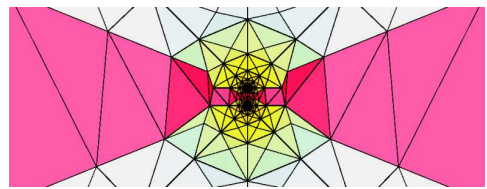}
  \caption{Initial molecular mesh for the CO molecule before
    refinement and improvement. The elements of the molecular region
    are shown in pink.}
  \label{fig:initial_molecular_mesh}
\end{figure}

The nodes for the molecular mesh region are then created by first
adding a spherical layer of nodes around the center of atomic charges
$\Rvec_{cc}$. The layer forms the boundary of the simulation cell and
has a radius equal to $r_{\partial \Omega} = q \max_i |\rvec_i -
\Rvec_{cc}|$, the radius of the furthest node from the center of
atomic charged multiplied by the layer ratio $q$. Then an
initial molecular mesh is created by a Delaunay tetrahedralization
\cite{bern_eppstein_95} of the nodes
(see Fig. \ref{fig:initial_molecular_mesh}).
The molecular mesh is then refined by Delaunay refinement
\cite{shewchuk_phd_97}, {\em i.e.}, by inserting nodes at the circumcenters
(the center of circumsphere) of too large elements one at the time and
repeating Delaunay tetrahedralization after each insertion. An element
is deemed too large, if
its longest edge is longer than the longest edge of an element in the
atomic mesh with the same distance from the closest atom. Or, if its
average edge length is longer than
the average edge length of an element in the atomic mesh with the same
distance from the closest atom. (Obviously, the elements, which are
connected to the nuclei, are ignored.)
After refining the mesh to fill the size constraints, the quality of
the elements is ensured. All elements with a too small ratio $s =
\sqrt{3 r_{in} / r_{circ}}$, where $r_{in}$ is the radius of the
inscribing sphere, and $r_{circ}$ is the radius of the circumsphere,
are Delaunay refined as above until no elements with low quality are
present. Keeping the ratio $s$ relatively close to one will ensure
that all angles (dihedral and face) are neither too large nor too small
\cite{shewchuk_good_fe_pdf, cavendish_85}. This is one of the standard
measures for the quality of an element. The elements which are
connected to the boundary nodes are not currently being refined.
However, the quality of these elements is not very important because
the solution is practically zero in this region.

\begin{figure}[ht]
  \includegraphics[width=80mm]{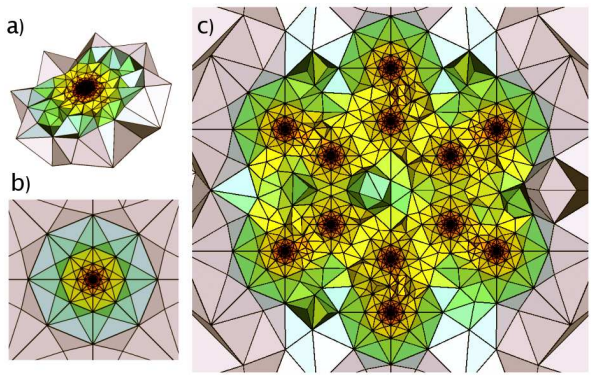}
  \caption{Cut plane of the molecular mesh of the C$_6$H$_6$ molecule
    with parameters $q=\sqrt{2}$, $s \geq \sqrt{1/3}$, and 15 outer
    layers (see text):
    a) the complete mesh (diameter of 55~\AA),
    b) the atomic mesh near a carbon nucleus,
    and c) the close-up of the molecular region.}
  \label{fig:c6h6_mesh}
\end{figure}

The resulting molecular mesh is somewhat finer than the atomic meshes,
but because the main interest is in the molecular region, we consider it
justified to slightly overdiscretize this region. An example of a
molecular mesh for benzene C$_6$H$_6$ with $q=\sqrt{2}$,
$s=\sqrt{1/3}$, and 15 outer layers is shown in Fig.
\ref{fig:c6h6_mesh}. The diameter of mesh is 55~\AA .

\subsection{Implementation}

Our current implementation is based on the ELMER finite element software
package \cite{elmer}, and the Delaunay tetrahedralization is done
using TETGEN \cite{tetgen, edelsbrunner_96}.
The ground-state Kohn-Sham system was solved with the self-consistent
iteration scheme. The locally optimal block preconditioned conjugated
gradient (LOBPCG) \cite{knyazev_lobpcg_01} method
was applied to the linearized Kohn-Sham eigenvalue problem
(Eq. \eqref{eq:Kohn_Sham_weak}), and the convergence rate of the
nonlinear system was enhanced with the Pulay mixing \cite{pulay_diis_80}
procedure for the density. The electronic charge was compensated by
Gaussian counter-charges at nuclei in the Poisson equation
(Eq. \eqref{eq:Poisson_weak}), and then a cancelling potential for the
counter-charges was added in the assembly of the Hamiltonian matrix
(in Eq. \eqref{eq:Kohn_Sham_fem}). Preconditioner for the eigenvalue
problem was chosen to be the incomplete Cholesky factorization
\cite{golub_ic} for $T + \alpha S$, where $T$ is the
kinetic energy operator and $\alpha$ was chosen to be $13.6$~eV.

In the linear-response calculation, the main effort is in calculating
the integrals of the matrix elements in equation
\eqref{eq:LR_integral}. Each row of the matrix is independent
of the other rows, and thus the problem is trivial to parallelize over
the rows of the matrix. Also, some of the matrix elements (and rows)
can be ignored beforehand as their eigenvalue difference is clearly
outside the relevant energy interval, {\em e.g.}, transitions from core
states.
The exchange-correlation kernel $f_{xc}(\rvec,\rvec',\omega)$ requires
the second functional derivative of the exchange-correlation functional
with respect to the density. However, when the second derivative is
not available, the finite-difference approximation
\begin{multline}
  \int d^3 \rvec \frac{\delta E_{xc}}{n(\rvec) n(\rvec')}
  n_{jk}(\rvec') =
  \\
  \lim_{\Delta \rightarrow 0}
  \frac{
    v_{xc}[n + \Delta n_{jk}](\rvec) - v_{xc}[n - \Delta n_{jk}](\rvec)
  }{ 2 \Delta }
\end{multline}
can be used. Above,
\begin{equation}
n_{jk}(\rvec) = \psi_j^*(\rvec) \psi_k(\rvec)
\end{equation}
is the pair density.

\section{Results and discussion}\label{sec:results}

We demonstrate our ground state DFT and linear response TDDFT methods
by applying them to atoms and small molecules. We calculated hydrogen,
carbon, and oxygen atoms, and hydrogen, carbon monoxide, and benzene
molecules. We calculated optical absorption spectra for a beryllium
atom, sodium dimer, and benzene molecule. The convergence
properties are discussed in both  cases.

\subsection{Ground state DFT}

We applied the local density approximation (LDA) functional with the
Perdew-Wang parametrization \cite{perdew_wang_92}
in all calculations, and all results are for spin-compensated
systems. In all calculations, the simulation cell diameter was
approximately 50~\AA, and the geometrical coarsening factor $q =
\sqrt{2}$.

The total energies of the atoms and molecules
calculated with increasing polynomial degree are shown in Tables
\ref{tbl:energies_atoms} and \ref{tbl:energies_molecules}, and
the atomization energies of the molecules in Table
\ref{tbl:atomization_molecules}. We have used for H$_2$ and CO the 
bond lengths of 0.75{\AA} and 1.1{\AA}, respectively. C$_6$H$_6$ has 
a planar geometry with atomic positions of
C:~$(0.000,\pm 1.396)${\AA}, $(\pm 1.209,\pm 0.698)${\AA}, and
H:~$(0.000, \pm 2.479)${\AA}, $(\pm 2.147, \pm 1.240)${\AA} used. The
H$_2$ mesh had 12$\times10^3$, 41$\times10^3$, and 96$\times10^3$
degrees of freedom (DOFs); the CO mesh had 14$\times10^3$,
46$\times10^3$, and 109$\times10^3$ DOFs; and the C$_6$H$_6$ mesh had
59$\times10^3$, 199$\times10^3$, and 470$\times10^3$ (DOFs), for
element degrees $p=2$, $p=3$, and $p=4$, respectively.
The corresponding results calculated with very high accuracy
($\sim$1~meV) using the electronic structure program FHI-aims
\cite{FHIaims} are shown on the last rows of the tables. As one can see, 
the total energy requires a high polynomial degree ($p>3$) to converge 
within an error below 100~meV. However, in practice one is interested
in the atomization energy of the system, which is the difference of the 
total energies between the system and the corresponding isolated atoms. 
The cancellation of errors leads to a significant improvement in 
the accuracy, and already the $2^{nd}$ and $3^{rd}$ degree
polynomials produce results with errors around 100~meV and 10~meV,
respectively. The maximal cancellation was obtained by using the same
mesh for isolated atoms as for the molecule, which can be considered
as a kind of a basis set superposition error, ({\em i.e.},
a counterpoise) correction \cite{boys_bsse_70}. The energies of the
isolated atoms are lower in the molecular mesh than in the atomistic
mesh. This is because the molecular mesh is denser than the atomistic
mesh as one wants to guarantee the good description of the bonding
regions. The total and atomization energies are well converged with
respect to the simulation cell diameter. We found less than one meV
difference in range from 21{\AA} to 151{\AA} for the CO molecule.

We performed nonrelativistic calculations for elements Zn, I, Hg, and
At in order to test the quality of the discretization in the case of
heavy elements. We found that elements with d-electrons perform
relatively well, e.g., the atomization energy of the I$_2$ molecule
($-2.400$eV, $-3.015$eV, $-3.031$eV for $p=2,3,4,$ respectively, and
$-3.037$eV for FHI-aims) has $\sim$2-4 times larger errors than the
C$_6$H$_6$ molecule. Elements with f-electrons perform much worse,
e.g., At$_2$ has on order of magnitude larger errors than C$_6$H$_6$
molecule. This is due to insufficient angular degrees of freedom as
the eigenvalues of the f-orbitals split (and d-orbitals split
slightly) in energy whereas p-orbitals do not. Our estimate is that
one would need $\sim$2-4 times more angular DOFs for heavy elements,
which in addition to $\sim$50\% more radial DOFs is $\sim$3-6 times
more DOFs than for carbon.

\begin{center}
\begin{table}[ht]
  \caption{Total energies of H, C, and O atoms calculated using
    elements with degrees $p=2-4$.}
  \label{tbl:energies_atoms}
  \begin{tabular}{l @{\hspace{5mm}}  c @{\hspace{5mm}} c @{\hspace{5mm}} c}
              & \multicolumn{3}{c}{$E_{LDA}$ [eV]} \\
              &       H  &         C  &         O  \\
    \hline
    p = 2     & -12.0509 & -1011.1067 & -2011.1970 \\
    p = 3     & -12.1245 & -1018.1042 & -2025.8759 \\
    p = 4     & -12.1271 & -1018.3581 & -2026.4268 \\
    FHI-aims  & -12.127 & -1018.369 & -2026.451 \\
  \end{tabular}
\end{table}
\end{center}

\begin{center}
\begin{table}[ht]
  \caption{Total energies of H$_2$, CO and C$_6$H$_6$ molecules
    calculated using elements with degrees $p=2-4$.}
  \label{tbl:energies_molecules}
  \begin{tabular}{l @{\hspace{5mm}}  c @{\hspace{5mm}} c @{\hspace{5mm}} c}
              & \multicolumn{3}{c}{$E_{LDA}$ [eV]}  \\
              &   H$_2$  &        CO  &  C$_6$H$_6$ \\
    \hline
    p = 2     & -30.8407 & -3039.5322 & -6226.5746 \\
    p = 3     & -30.9510 & -3059.7776 & -6262.5718 \\
    p = 4     & -30.9542 & -3060.5009 & -6263.7841 \\
    FHI-aims  & -30.954 & -3060.529 & -6263.829 \\
  \end{tabular}
\end{table}
\end{center}

\begin{center}
\begin{table}[ht]
  \caption{Atomization energies of H$_2$, CO and C$_6$H$_6$ molecules
    calculated using elements with degrees $p=2-4$.}
  \label{tbl:atomization_molecules}
  \begin{tabular}{l @{\hspace{5mm}}  c @{\hspace{5mm}} c @{\hspace{5mm}} c}
              & \multicolumn{3}{c}{$\Delta E_{LDA}$ [eV]}  \\
              &   H$_2$ &      CO  &  C$_6$H$_6$ \\
    \hline
    p = 2     & -6.6838 & -15.7573 & -81.0894 \\
    p = 3     & -6.6996 & -15.7162 & -80.8599 \\
    p = 4     & -6.6999 & -15.7114 & -80.8541 \\
    FHI-aims  & -6.700 & -15.709 & -80.852 \\
  \end{tabular}
\end{table}
\end{center}

Tables \ref{tbl:CO_atomization} and \ref{tbl:CO_dm} show the
convergence of the potential energy surface and the dipole moment,
respectively, calculated with elements with degrees $p=2-4$. The
potential energy surface shows no ``egg-box effect'', known to exists
in uniform real-space grids \cite{ono_double_grid_99}. However, there
exists a similar kind of effect. For example in a diatomic molecule,
when the bond length is changed, new elements are created into or old
ones are removed from the mesh. In improperly generated meshes, this
can cause severe problems as the potential energy surface may have
significant artificial oscillations and discontinuities. For this
reason, we recommend a slightly denser discretization of the bonding
regions compared to the atomic regions. Based on our experimentations
on diatomic molecules, this is sufficient and forces with a quality
comparable to that from commonly used codes, such as the real-space
code GPAW \cite{GPAW}, are obtained.

Note, that we have given two different values for the atomization
energy of CO at the bond length of $R_{CO} = 1.1 \text{\AA}$ for each
element degree $p$ (see Tables \ref{tbl:atomization_molecules} and
\ref{tbl:CO_atomization}). Because the mesh generation
is not unique for a given molecule but rather for given Cartesian
positions and the order in which the atoms are given, the
difference is due to different meshes obtained from two different
generator inputs. However, the difference is one order of magnitude
smaller than the error in the atomization energy. The dipole moment
shows errors less than 0.01~$e${\AA} and 0.001~$e${\AA} when using 2nd
and 3rd order polynomials, respectively.

\begin{table}[ht]
  \caption{Atomization energy of the CO molecule at different bond lengths
    calculated using elements with degrees $p=2-4$.}
  \label{tbl:CO_atomization}
  \begin{tabular}{c @{\hspace{3mm}} r @{\hspace{5mm}} r
      @{\hspace{5mm}} r @{\hspace{5mm}} r}
         & \multicolumn{4}{c}{$\Delta E_{LDA}$ [eV]} \\
    $R_{CO} [\text{\AA}]$&  p = 2  &  p = 3  &  p = 4  & FHI-aims \\
    \hline
     0.8 & -0.1272 & -0.6514 & -0.6648 &  -0.660 \\
     1.0 &-14.4446 &-14.4495 &-14.4464 & -14.444 \\
     1.1 &-15.7584 &-15.7175 &-15.7115 & -15.709 \\
     1.2 &-15.6235 &-15.4910 &-15.4845 & -15.482 \\
     1.4 &-13.5165 &-13.3027 &-13.2934 & -13.292 \\
     1.8 & -8.5848 & -8.3963 & -8.3875 &  -8.386 \\
     2.4 & -4.0303 & -3.9093 & -3.9043 &  -3.903 \\
  \end{tabular}
\end{table}

\begin{table}[ht]
  \caption{Dipole moment of the CO molecule at different bond lengths
    calculated using elements with degrees $p=2-4$.}
  \label{tbl:CO_dm}
  \begin{tabular}{c @{\hspace{3mm}} r @{\hspace{5mm}} r
      @{\hspace{5mm}} r @{\hspace{5mm}} r}
           & \multicolumn{4}{c}{$\mu_{LDA} $ [$e\text{\AA}$]} \\
    $R_{CO}[\text{\AA}]$&  p = 2  &  p = 3 &  p = 4  & FHI-aims \\
    \hline
     0.8   & 0.2454 & 0.2402 & 0.2400 & 0.2398 \\
     1.0   & 0.1390 & 0.1311 & 0.1307 & 0.1305 \\
     1.1   & 0.0745 & 0.0669 & 0.0666 & 0.0663 \\
     1.2   & 0.0064 &-0.0010 &-0.0013 &-0.0015 \\
     1.4   &-0.1330 &-0.1397 &-0.1399 &-0.1399 \\
     1.8   &-0.3792 &-0.3792 &-0.3791 &-0.3790 \\
     2.4   &-0.6084 &-0.5996 &-0.5992 &-0.5991 \\
  \end{tabular}
\end{table}

In Table \ref{tbl:C6H6_eigs}, we show the Kohn-Sham eigenvalues of the
C$_6$H$_6$ molecule. The core eigenvalues exhibit much larger absolute errors
than the valence eigenvalues, but the relative errors are of same order. The
valence eigenvalues converge similarly to the atomization energies,
which is reasonable as the errors in the core eigenvalues
cancel when taking the differences. The remaining error is mainly due
to the valence states and the molecular orbitals which they form.

\begin{table}[ht]
  \caption{Kohn-Sham orbital energies (eigenvalues) of the C$_6$H$_6$
    molecule calculated using elements with degrees $p=2-4$.}
  \label{tbl:C6H6_eigs}
  \begin{tabular}{c @{\hspace{5mm}} r @{\hspace{5mm}} r
      @{\hspace{5mm}} r @{\hspace{5mm}} r}
           & \multicolumn{4}{c}{$\epsilon_{LDA}$ [eV]} \\
     state &  p = 2  &  p = 3  &  p = 4    & FHI-aims \\
    \hline
     1 & -264.6616 & -266.3819 & -266.4388 & -266.4382 \\
     $\cdots$  & & & & \\
     6 & -264.6087 & -266.3585 & -266.4156 & -266.4150 \\
     7 &  -21.1552 &  -21.1155 &  -21.1557 &  -21.1560 \\
     8 &  -18.3474 &  -18.3608 &  -18.3616 &  -18.3619 \\
     9 &  -18.3404 &  -18.3597 &  -18.3609 &  -18.3612 \\
     $\cdots$  & & & & \\
    18 &   -8.2867 &   -8.2915 &   -8.2915 &   -8.2917 \\
    19 &   -8.2839 &   -8.2895 &   -8.2895 &   -8.2897 \\
    20 &   -6.5401 &   -6.5341 &   -6.5343 &   -6.5338 \\
    21 &   -6.5385 &   -6.5339 &   -6.5342 &   -6.5338 \\
  \end{tabular}
\end{table}

\subsection{Linear-response TDDFT}

For the linear-response TDDFT calculations we used actually a
slightly different mesh generation scheme than that described
above in Sec.~\ref{mesh_generation}. This old scheme,
developed also by us, uses i) different alternating polyhedra,
{\em i.e.} tetrakis hexahedron and slightly compressed
(larger cubic faces) truncated cuboctahedron, for atomic
meshes, and ii) different quality measures, {\em i.e.} dihedral angles
and aspect ratio (longest edge / smallest side height), than the
current one. Compared to the old one, the current mesh generation
scheme is simpler and it produces higher quality atomic
meshes. However, the difference in quality is negligible when applying
to the linear-response TDDFT.

\begin{figure}[ht]
  \includegraphics{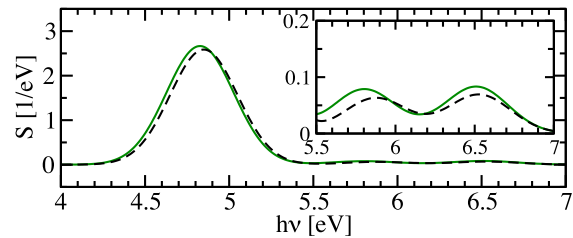}
  \caption{Optical absorption spectra of the beryllium atom
    calculated using elements with degrees $p=2$ (solid) and $p=3$
    (dashed). The inset shows a magnification of the high-energy region.}
  \label{fig:be_p2_p3_spectrum}
\end{figure}

\begin{figure}[ht]
  \includegraphics{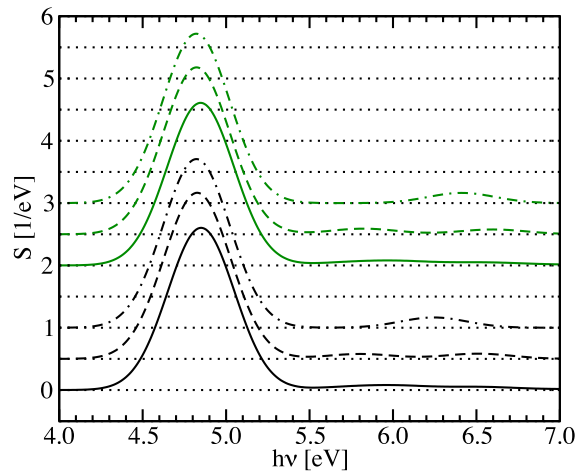}
  \caption{Optical absorption spectrum of the beryllium atom calculated
    using the confinement potential parameters (from the highest curve to 
    the lowest one):
    $k_c = 10^{-2} E_h/a_0^2$, $R_c = 4.0 a_0$;
    $k_c = 10^{-3} E_h/a_0^2$, $R_c = 4.0 a_0$;
    $k_c = 10^{-4} E_h/a_0^2$, $R_c = 4.0 a_0$;
    $k_c = 10^{-2} E_h/a_0^2$, $R_c = 8.0 a_0$;
    $k_c = 10^{-3} E_h/a_0^2$, $R_c = 8.0 a_0$; and
    $k_c = 10^{-4} E_h/a_0^2$, $R_c = 8.0 a_0$.
    The spectra are separated by shifting the zero level.
  }
  \label{fig:be_p2_conf_spectrum}
\end{figure}

\begin{figure}[ht]
  \includegraphics{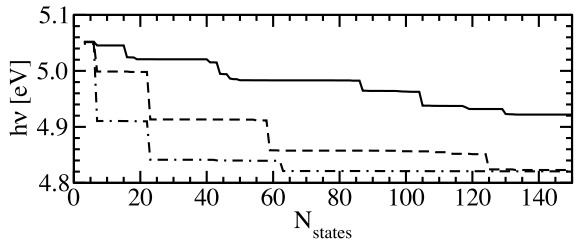}
  \caption{Convergence of the position of the first transition peak 
    in the optical absorption spectrum of the beryllium atom with 
    respect to the number of states included in the calculation. 
    The confinement potential parameters used are:
    $k_c = 10^{-2} E_h/a_0^2$, $R_c = 4.0 a_0$ (dash-dotted);
    $k_c = 10^{-3} E_h/a_0^2$, $R_c = 4.0 a_0$ (dashed); and
    $k_c = 10^{-4} E_h/a_0^2$, $R_c = 4.0 a_0$ (solid).}
  \label{fig:be_p2_first_trans_conv}
\end{figure}

First, we consider a simple test system, a beryllium atom, to
demonstrate the convergence properties. We begin with the polynomial degrees
$p=2$ and $p=3$, 150 states, the confinement radius $R_c=8.0 a_0$ and the
force constant $k_c = 10^{-3} E_h/a_0^2$. The resulting spectra are
shown in Fig. \ref{fig:be_p2_p3_spectrum}. Increasing the polynomial
degree of the elements has only a small effect of $\sim$20~meV for the
first peak position, and of $\sim$70~meV for the second peak position
($h\nu_{p=3} > h\nu_{p=2}$). The effect of different confinement
potentials can be seen in Figs. \ref{fig:be_p2_conf_spectrum} and
\ref{fig:be_p2_first_trans_conv}. A stronger confinement provides
a faster convergence with respect to the number of states, but at the same
time, the converged transition energies are shifted to slightly higher
energies. A weaker confinement provides energies which are better converged, 
but the convergence may not be reached with the
available number of states, as in the case of $k_c = 10^{-4}
E_h/a_0^2$ in Fig. \ref{fig:be_p2_first_trans_conv}. In Fig.
\ref{fig:be_p2_conf_spectrum}, the number of states was increased to
250 which yields an error less than 30~meVs. Obviously, the
transitions at higher energies are more sensitive to confinement than
transitions at low energies. The convergence with respect to the
number of states included in the calculation is not smooth, but rather
has a step every time a new state contributing to the transition is
included in the basis. The step is not always smaller than the
previous one, and it can be hard to decide whether the spectrum has
converged by observing the convergence with respect to the number of
states.

\begin{figure}[ht]
  \includegraphics{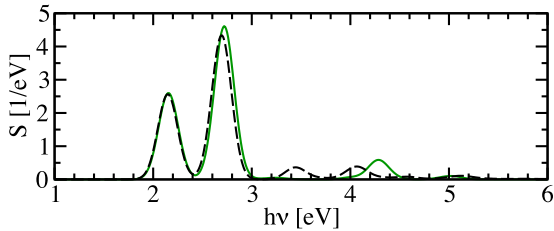}
  \caption{Optical absorption spectra of the sodium dimer calculated with
    the confinement potential parameters of
    $k_c = 10^{-2} E_h/a_0^2$, $R_c = 8.0 a_0$ (solid); and
    $k_c = 10^{-3} E_h/a_0^2$, $R_c = 8.0 a_0$ (dashed).}
  \label{fig:na2_spectrum}
\end{figure}

Next, we examined two molecular test systems, the sodium dimer Na$_2$
and the benzene molecule C$_6$H$_6$. The simulated photoabsorption
spectrum of the Na$_2$ is shown in Fig. \ref{fig:na2_spectrum}. The
calculation included 250 states, and two different confinement
potentials were used: one with $R_c = 8.0 a_0$ and $k_c = 10^{-2}
E_h/a_0^2$, and one with $R_c = 8.0 a_0$ and $k_c = 10^{-3}
E_h/a_0^2$. Practically, the same result of 2.15~eV was obtained for
the first peak wit the two sets of parameters. For the second one 
there is a small shift from
2.69~eV to 2.72~eV. In contrast, the third clearly visible peak in the
spectrum shows a remarkable shift from 3.4~eV to 4.3~eV.

\begin{figure}[ht]
    \includegraphics{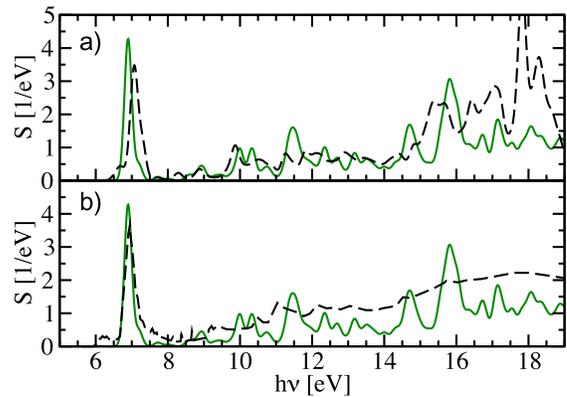}
  \caption{Optical absorption spectra of the benzene molecule:
    a) the spectra calculated using the confinement potential parameters
    $k_c = 10^{-2} E_h/a_0^2$, $R_c = 4.0 a_0$ (solid); and
    $k_c = 10^{-3} E_h/a_0^2$, $R_c = 4.0 a_0$ (dashed),
    b) the spectrum calculated using the confinement potential parameters
    $k_c = 10^{-2} E_h/a_0^2$, $R_c = 4.0 a_0$ (solid) and the
    experimental spectrum \cite{koch_c6h6_exper_72} (dashed).}
  \label{fig:c6h6_spectrum}
\end{figure}

The photoabsorption spectrum of the benzene molecule is shown in
Fig. \ref{fig:c6h6_spectrum}. Again two different confinement
potentials were used, one with $R_c = 4.0 a_0$ and $k_c = 10^{-2}
E_h/a_0^2$, and one with $R_c = 4.0 a_0$ and $k_c = 10^{-3}
E_h/a_0^2$. The spectrum with the weaker confinement  ($k_c =
10^{-3}$) is not converged yet with 250 states, which corresponds
already nearly 4 million matrix elements. The spectrum with the
stronger confinement and 150 states is converged in the lower energy
part of the spectrum, and reproduces correctly the main experimental peak 
around 7~eV. It also shows the beginning of a broad feature above
9~eV in agreement with the experiment.

\subsection{Computational details}

The ground state DFT calculations were performed as serial
calculations, and the time consumed ranged from minutes (hydrogen
atom) to tens of hours (benzene with $p = 4$). All calculations were
done on 2.6~GHz AMD Opteron dual-core processors. As the systems were
relatively small, the storage requirements of the matrices were much
larger than those of the wavefunctions. The number of
nonzero entries in the matrices ranged from $1\times 10^5$ (H, $p=2$) to
$4\times 10^7$ (C$_6$H$_6$, $p=4$). The number of degrees of freedom
ranged from $5000$ (H, $p=2$) to $5\times 10^5$ (C$_6$H$_6$, $p=4$).
The linear response TDDFT was parallelized over the rows of the
Casida matrix, and the absorption spectrum of benzene was calculated
using several hundreds of processors.

We consider the performance attained adequate for an initial
``proof-of-concept'' implementation. And, we expect to increase the speed
substantially by employing more sophisticated methods. Especially, the
preconditioning of the eigenvalue problem and improved initial guesses
for Kohn-Sham wavefunctions are expected to result in remarkable
improvements.

\section{Conclusions}

We have described and implemented a high-order hierarchical finite element
method on unstructured meshes for all-electron DFT and TDDFT
method. Our finite element mesh generation scheme assures the quality
of the elements in the mesh by merging high-quality, structured atomic
meshes to an initial molecular mesh, which is then refined to meet
the size and shape requirements by applying the Delaunay refinement
method. The ground state DFT calculations were performed using
elements with degrees $p = 2-4$, which provide increasing levels of accuracy
down to a few meVs.

We also described a flexible way to construct a basis for the
finite-element linear response TDDFT calculation. By applying an
auxiliary confinement potential to the ground-state calculation, the
basis can be tuned to balance between accuracy and computational
cost. The convergence properties of the optical absorption spectrum
were discussed in the cases of the beryllium atom, and the sodium
dimer and benzene molecules.

The initial implementation has proved the applicability of the
hierarchical finite element method on unstructured meshes to all-electron
DFT and TDDFT. However, there exist several open question, which
must be further studied and improved, for example, the preconditioning
of the eigenvalue problem. As the finite element method is well-suited for
the domain decomposition, the parallel implementation would provide
access to much larger systems within reasonable execution times. As
most of the applications do not need full all-electron solutions, the PAW
method or a similar treatment should speed up calculations remarkably in
these cases. Magnetic fields, relativistic effects, and quantum
mechanical forces for atoms will be implemented in order to
broaden the applicability of the method. Finally, we believe that the most
promising application areas for our method are beyond the ground-state
and linear response calculations, for example, in the time-propagation
TDDFT scheme.

\section{Acknowledgements}

This work was performed in COMP which is Center of Excellence of
Academy of Finland, and was funded by a grant from CSC - IT Center for
Science Ltd., AMD, and Cray. The calculations were performed using
CSC's resources. L.L. wishes to thank Mikko Lyly and the rest of CSC's
Elmer team for support with Elmer, as well as Harri Hakula for
advices on mesh generation. Elmer is an open source finite element
software for multiphysical problems developed and supported by CSC.

\bibliographystyle{plain}
\bibliography{all_electron}

\begin{thebibliography}{10}

\bibitem{batcho_98}
P.F. Batcho.
\newblock Spectrally accurate numerical solution of the single-particle schr\"o
  dinger equation.
\newblock {\em Phys. Rev. A}, 57:4246--4252, 1998.

\bibitem{batcho_00}
P.F. Batcho.
\newblock Computational method for general multicenter electronic structure
  calculations.
\newblock {\em Phys. Rev. E}, 61:7169--7183, 2000.

\bibitem{bathe_fem_book}
K.-J. Bathe.
\newblock {\em Finite Element Procedures (Part 1-2), 2nd ed.}
\newblock Prentice Hall, 1995.

\bibitem{bern_eppstein_95}
M.~Bern and D.~Eppstein.
\newblock Mesh generation and optimal triangulation.
\newblock In D.-Z. Du and F.K. Hwang, editors, {\em Computing in Euclidean
  Geometry, 2nd ed.}, pages 47--123. World Scientific, 1995.

\bibitem{bloch_paw_94}
P.~E. Bl\"ochl.
\newblock Projector augmented-wave method.
\newblock {\em Phys. Rev. B}, 50:17953, 1994.

\bibitem{FHIaims}
V.~Blum, R.~Gehrke, F.~Hanke, P.~Havu, V.~Havu, X.~Ren, K.~Reuter, and
  M.~Scheffler.
\newblock \emph{Ab initio} molecular simulations with numeric atom-centered
  orbitals: {FHI-aims}.
\newblock {\em Comp. Phys. Comm.}, 2008.
\newblock Accepted for publication.

\bibitem{boys_bsse_70}
S.~F. Boys and F.~Bernardi.
\newblock The calculation of small molecular interactions by the differences of
  separate total energies. some procedures with reduced errors.
\newblock {\em Molecular Physics}, 19:553--566, 1970.

\bibitem{brenner_scott_book}
S.~C. Brenner and L.~R. Scott.
\newblock {\em The Mathematical Theory of Finite Element Methods}.
\newblock Springer, 2008.

\bibitem{casida_lrtddft_96}
M.~E. Casida.
\newblock Time-dependent density functional response theory of molecular
  systems: Theory, computational methods, and functionals.
\newblock In J.~M. Seminario, editor, {\em Recent Developments and Applications
  in Modern Density-Functional Theory}, page 391. Elsevier, Amsterdam, 1996.

\bibitem{casida_lrtddft_95}
M.E. Casida.
\newblock Time-dependent density-functional response theory for molecules.
\newblock In D.P. Chong, editor, {\em Recent Advances in Density Functional
  Methods, Part I}, page 155. World Scientific, Singapore, 1995.

\bibitem{cavendish_85}
J.~C. Cavendish, D.~A. Field, and W.~H. Frey.
\newblock An apporach to automatic three-dimensional finite element mesh
  generation.
\newblock {\em International Journal for Numerical Methods in Engineering},
  21:329--347, 1985.

\bibitem{edelsbrunner_96}
H.~Edelsbrunner and N.~R. Shah.
\newblock Incremental topological flipping works for regular triangulations.
\newblock {\em Algorithmica}, 15:223--241, 1996.

\bibitem{elliott_tddft_review_2008}
P.~Elliott, F.~Furche, and K.~Burke.
\newblock Excited states from time-dependent density functional theory.
\newblock {\em Rev. Comp. Chem.}, 26:91--165, 2008.

\bibitem{fattebert_2007}
J.-L. Fattebert, R.~D. Hornunga, and A.~M. Wissink.
\newblock Finite element approach for density functional theory calculations on
  locally-refined meshes.
\newblock {\em J. Comp. Phys}, 223:759--773, 2007.

\bibitem{fermi_pp_34}
E.~Fermi.
\newblock Sopra lo spostamento per pressione delle righe elevate delle serie
  spettrali.
\newblock {\em Nuovo Cimento}, 11:157, 1934.

\bibitem{dft_primer}
C.~Fiolhais, F.~Nogueira, and M.~Marques, editors.
\newblock {\em A Primer in Density Functional Theory}.
\newblock Springer-Verlag, 2003.

\bibitem{elmer}
CSC IT~Center for Science.
\newblock Elmer, open source finite element software for multiphysical
  problems.
\newblock http://www.csc.fi/english/pages/elmer.

\bibitem{Gaussian03}
M.~J. {Frisch {\em et al.}}
\newblock Gaussian 03.
\newblock \uppercase{G}aussian, Inc., Wallingford, CT, 2004.

\bibitem{golub_ic}
G.~H. Golub and C.~F. {Van Loan}.
\newblock {\em Matrix Computations, 2nd ed.}, pages 530--531.
\newblock Johns Hopkins, 1993.

\bibitem{AbInit}
X.~Gonze, J.-M. Beuken, R.~Caracas, F.~Detraux, M.~Fuchs, G.-M. Rignanese,
  L.~Sindic, M.~Verstraete, G.~Zerah, F.~Jollet, M.~Torrent, A.~Roy, M.~Mikami,
  Ph. Ghosez, J.-Y. Raty, and D.C. Allan.
\newblock First-principles computation of material properties : the abinit
  software project.
\newblock {\em Comp. Mat. Sci.}, 25:478--492, 2002.

\bibitem{hohenberg_kohn_64}
P.~Hohenberg and W.~Kohn.
\newblock Inhomogeneous electron gas.
\newblock {\em Phys. Rev.}, 136:B864--B871, 1964.

\bibitem{knyazev_lobpcg_01}
A.~Knyazev.
\newblock Toward the optimal preconditioned eigensolver: Locally optimal block
  preconditioned conjugate gradient method.
\newblock {\em SIAM Journal on Scientific Computing}, 23:517--541, 2001.

\bibitem{koch_c6h6_exper_72}
E.~E. Koch and A.~Otto.
\newblock Optical absorption of benzene vapour for photon energies from 6 ev to
  35 ev.
\newblock {\em Chem. Phys. Lett.}, 12:476--480, 1972.

\bibitem{kohn_nobel_99}
W.~Kohn.
\newblock Nobel lecture: Electronic structure of matter -- wave functions and
  density functionals.
\newblock {\em Rev. Mod. Phys.}, 71:1253--1266, 1999.

\bibitem{kohn_sham_65}
W.~Kohn and L.~J. Sham.
\newblock Self-consistent equations including exchange and correlation effects.
\newblock {\em Phys. Rev.}, 140:A1133--A1138, 1965.

\bibitem{VASP}
G.~Kresse and J.~Hafner.
\newblock Ab initio molecular dynamics for liquid metals.
\newblock {\em Phys. Rev. B}, 47:558, 1993.

\bibitem{levin_85}
F.S. Levin and J.~Shertzer.
\newblock Finite-element solution of the schr\"ondiger equation for the helium
  ground state.
\newblock {\em Phys. Rev. A}, 32:3285, 1985.

\bibitem{Octopus}
M.~A.~L. Marques, A.~Castro, G.~F. Bertsch, and A.~Rubio.
\newblock octopus: a first-principles tool for excited electron-ion dynamics.
\newblock {\em Comp. Phys. Comm.}, 151:60--78, 2003.

\bibitem{GPAW}
J.~J. Mortensen, L.~B. Hansen, and K.~W. Jacobsen.
\newblock Real-space grid implementation of the projector augmented wave
  method.
\newblock {\em Phys. Rev. B}, 71:035109, 2005.

\bibitem{ono_double_grid_99}
T.~Ono and K.~Hirose.
\newblock Timesaving double-grid method for real-space electronic-structure
  calculations.
\newblock {\em Phys. Rev. Lett.}, 82:5016--5019, 1999.

\bibitem{parr_yang_dft_book}
R.~G. Parr and W.~Yang.
\newblock {\em Density-Functional Theory of Atoms and Molecules}.
\newblock Oxford University Press, 1989.

\bibitem{pask_sterne_review_05}
J.~E. Pask and P.~A. Sterne.
\newblock Finite element methods in ab initio electronic structure
  calculations.
\newblock {\em Modelling Simul. Matter. Sci. Eng.}, 13:R71--R96, 2005.

\bibitem{perdew_wang_92}
J.~P. Perdew and Y.~Wang.
\newblock Accurate and simple analytic representation of the electron-gas
  correlation energy.
\newblock {\em Phys. Rev. B}, 45:13244--13249, 1992.

\bibitem{phillips_pp_59}
J.~C. Phillips and L.~Kleinman.
\newblock New method for calculating wave functions in crystals and molecules.
\newblock {\em Phys. Rev.}, 116:287--294, 1959.

\bibitem{pulay_diis_80}
P.~Pulay.
\newblock Convergence acceleration of iterative sequences. {T}he case of {SCF}
  iteration.
\newblock {\em Chem. Phys. Lett.}, 73:393--398, 1980.

\bibitem{runge_gross_84}
E.~Runge and E.~K.~U. Gross.
\newblock Density-functional theory for time-dependent systems.
\newblock {\em Phys. Rev. Lett.}, 52:997--1000, 1984.

\bibitem{schwab_hp_book_98}
Ch. Schwab.
\newblock {\em {$p$}- and {$hp$}-finite element methods}.
\newblock The Clarendon Press Oxford University Press, New York, 1998.

\bibitem{shewchuk_phd_97}
J.~R. Shewchuk.
\newblock {\em Delaunay Refinement Mesh Generation}.
\newblock PhD thesis, School of Computer Science, Carnegie Mellon University,
  Pittsburgh, Pennsylvania, May 1997.
\newblock Available as Technical Report CMU-CS-97-137.

\bibitem{shewchuk_good_fe_pdf}
J.~R. Shewchuk.
\newblock What is a good linear finite element? interpolation, conditioning,
  anisotropy, and quality measures.
\newblock http://www.cs.berkeley.edu/~jrs/papers/elemj.pdf, 2002.
\newblock unpublished preprint.

\bibitem{tetgen}
Hang Si.
\newblock Tetgen, three-dimensional delaunay triangulator.
\newblock http://tetgen.berlios.de/.

\bibitem{Siesta}
J.~M. Soler, E.~Artacho, J.~D. Gale, A.~García, J.~Junquera, P.~Ordejon, and
  D.~Sanchez-Portal.
\newblock The siesta method for ab initio order-n materials simulation.
\newblock {\em J. Phys.: Cond Matter}, 14:2745--2779, 2002.

\bibitem{szabo_babuska_fea_book}
B.~Szabo and I.~Babuska.
\newblock {\em Finite Element Analysis}.
\newblock Wiley Interscience, 1991.

\bibitem{troullier_martins_pp_91}
N.~Troullier and Jose~Luis Martins.
\newblock Efficient pseudopotentials for plane-wave calculations.
\newblock {\em Phys. Rev. B}, 43:1993--2206, 1991.

\bibitem{tsuchida_95}
E.~Tsuchida and M.~Tsukada.
\newblock Electronic-structure calculations based on the finite-element method.
\newblock {\em Phys. Rev. B}, 52:5573, 1995.

\bibitem{turner_fem_56}
M.~J. Turner, R.~W. Clough, H.~C. Martin, and L.~J. Topp.
\newblock Stiffness and deflection analysis of complex structures.
\newblock {\em Journal of Aeronautical Sciences}, 23:805--824, 1956.

\bibitem{white_89}
S.~R. White, J.~W. Wilkins, and M.~P. Teter.
\newblock Finite-element method for electronic structure.
\newblock {\em Phys. Rev. B}, 39:5819, 1989.

\bibitem{yabana_tdlda_96}
K.~Yabana and G.~F. Bertsch.
\newblock Time-dependent local-density approximation in real time.
\newblock {\em Phys. Rev. B}, 54:4484--4487, 1996.

\bibitem{yu_h2_fem_94}
Hengtai Yu and Andre~D. Bandrauk.
\newblock Three-dimensional cartesian finite element method for the time
  dependent schrödinger equation of molecules in laser fields.
\newblock {\em J. Chem. Phys.}, 102:1257--1265, 1994.

\bibitem{zhang_2008}
D.~Zhang, L.~Shen, A.~Zhou, and X.-G. Gong.
\newblock Finite element method for solving kohn-sham equations based on
  self-adaptive tetrahedral mesh.
\newblock {\em Phys. Lett. A}, 372:5071--5076, 2008.

\end{thebibliography}

\end{document}